\documentclass[twocolumn]{aastex63}

\usepackage{sidecap}
\usepackage{soul}

\definecolor{darkgreen}{RGB}{0,150,20}

\begin{document}

\title{Interaction of magnetic fields with a vortex tube
at solar subgranular scale}

\correspondingauthor{C.E.~Fischer}
\email{cfischer@leibniz-kis.de}

\author{C.E.~Fischer}
\affiliation{Leibniz Institute for Solar Physics (KIS)\\ Sch\"{o}neckstrasse 6, 79104 Freiburg, Germany}

\author{G.~Vigeesh}
\affiliation{Leibniz Institute for Solar Physics (KIS)\\ Sch\"{o}neckstrasse 6, 79104 Freiburg, Germany}

\author{P.~Lindner}
\affiliation{Leibniz Institute for Solar Physics (KIS)\\ Sch\"{o}neckstrasse 6, 79104 Freiburg, Germany}

\author{J.M.~Borrero}
\affiliation{Leibniz Institute for Solar Physics (KIS)\\ Sch\"{o}neckstrasse 6, 79104 Freiburg, Germany}

\author{F.~Calvo}
\affiliation{Institute for Solar Physics, Dept.\ of Astronomy, Stockholm University\\
 AlbaNova University Centre,
106 91 Stockholm, Sweden} 

\author{O.~Steiner}
\affiliation{Leibniz Institute for Solar Physics (KIS)\\ Sch\"{o}neckstrasse 6, 79104 Freiburg, Germany}
\affiliation{Istituto Ricerche Solari Locarno (IRSOL)\\ Via Patocchi 57,
6605 Locarno-Monti, Switzerland}

%



\begin{abstract}

Using high-resolution spectropolarimetric data recorded with the Swedish 1 m Solar Telescope, we have identified several instances of granular lanes traveling into granules. These are believed to be the observational signature of underlying  tubes of vortical flow with their axis oriented parallel to the solar surface. Associated with these horizontal vortex tubes, we detect in some cases a significant signal in linear polarization, located at the trailing dark edge of the granular lane. The linear polarization appears at a later stage of the granular lane development, and is flanked by patches of circular polarization. Stokes inversions show that the elongated patch of linear polarization signal arises from the horizontal magnetic field aligned with the granular lane. We analyze snapshots of a magnetohydrodynamic numerical simulation and find cases in which the horizontal  vortex tube of the granular lane  redistributes and transports the magnetic field to the solar surface causing a polarimetric signature similar to what is observed. We  thus witness a mechanism capable of  transporting magnetic flux to the solar surface within granules. This mechanism is probably an important component of the small-scale dynamo supposedly acting at the solar surface and generating the quiet-Sun magnetic field.

\end{abstract}

\keywords{Sun: photosphere --- Sun: granulation --- 
Sun: magnetic fields --- dynamo --- polarization --- magnetohydrodynamics (MHD)}

\section{Introduction} \label{sec:intro}

Granular lanes, first observed and termed so by \cite{2010ApJ...723L.180S} consist of a dark lane  preceded by a bright rim that both move together from the boundary of a granule into the granule itself.
The trailing dark lane often remains attached to the edges  of the parent granule giving it an arch-like appearance as can be seen, e.g., in the first row of Fig.~\ref{tempdevelop2}.~\cite{2010ApJ...723L.180S} performed further analysis by searching for this signature in continuum images synthesized from CO$^{5}$BOLD \citep{2012JCoPh.231..919F} simulations.  Examining 32 simulation cases, they concluded that the observed signature was caused by underlying horizontal vortex tubes aligned to the solar surface. Interestingly, a connection of the magnetic field with horizontal vortex tubes was found in some of the simulation cases but could not be confirmed in observations of granular lanes with the Imaging Magnetograph eXperiment \citep[IMaX;][]{2011SoPh..268...57M}.

Since then, horizontal vortex tubes in simulations have also been reported by others using different codes \citep{2012PhyS...86a8403K,2013arXiv1312.0982K,2018ApJ...859..161R}, which reproduced in each case the overall structure of a vortex tube beneath the observable intensity signal  from the solar surface. Horizontal vortex tubes seem to be the natural consequence of vorticity generated by the baroclinic effect (nonaligned density and pressure gradients) near the edges of granules.  
The observational counterpart ---\,the granular lane\,--- is a common occurrence and easily recognizable in high-resolution continuum time series from the largest solar telescopes.~\cite{2011ApJ...736L..35Y}  observed granular lanes in TiO filtergrams and speculated from simultaneously recorded H$\alpha - 0.1$\,nm and H$\alpha + 0.07$\,nm filtergrams that cospatially occurring chromospheric jets were caused by the interaction of magnetic fields associated with the granular lane with preexisting magnetic fields.~\cite{2012arXiv1207.6418Y} found in high-resolution polarimetric data from the Goode Solar Telescope~\citep{2010AN....331..620G} diffuse intensity structures within the granules exhibiting flux emergence. In some cases these intensity structures were clearly due to granular lanes, which led them to suggest that horizontal vortex tubes might in some way be related to small-scale magnetic field emergence. In this letter we  present polarimetric data from the Swedish 1 m Solar Telescope~\citep[SST;][]{2003SPIE.4853..341S}, which for the first time show cases of distinct and persistent polarization signal developing with granular lanes. Further insight is obtained by studying magnetohydrodynamic simulations of granular lanes with associated horizontal vortex tubes hinting at the local (shallow) recycling and generation of magnetic field associated with vortex tubes.

\section{Data and Methods} \label{sec:obs}
We studied a quiet-Sun disc-center full Stokes time series recorded with the CRisp Imaging SpectroPolarimeter ~\citep[CRISP;][]{2008ApJ...689L..69S} instrument attached to the SST. The data were obtained during a 9 day observing campaign in 2019 April as part of the SOLARNET access program. The data provided were processed using the SSTRED data pipeline~\citep[see][]{2018arXiv180403030L,2015A&A...573A..40D}, where it was corrected for dark current and a flat-field correction was applied. The data were further demodulated and corrected for cross talk (Stokes $I$ to $Q$, $U$, and $V$). A reconstruction was performed using the Multi-Object Multi-Frame Blind Deconvolution~\citep[MOMFBD;][]{2002SPIE.4792..146L,2005SoPh..228..191V} algorithm.
The observations were taken in the Fe\,{\sc{i}}\,617.3\,nm (Land\'{e}-factor $g_{\rm eff}=2.5$) spectral line. The line was sampled at 15 wavelength points spread symmetrically around the line core ($\lambda_{0}$) in 3.5\,pm steps.
CRISP has a field of view (FOV) of approximately $50\arcsec \times 50\arcsec$, and the pixel size is $0.\arcsec 059 \times 0.\arcsec 059$. The cadence of the data was approximately 28\,s. A mean seeing parameter $r_{0}$ of 22.7\,cm was recorded during the 2 hr time sequence. 

\begin{figure*}
  \includegraphics[width=\textwidth]{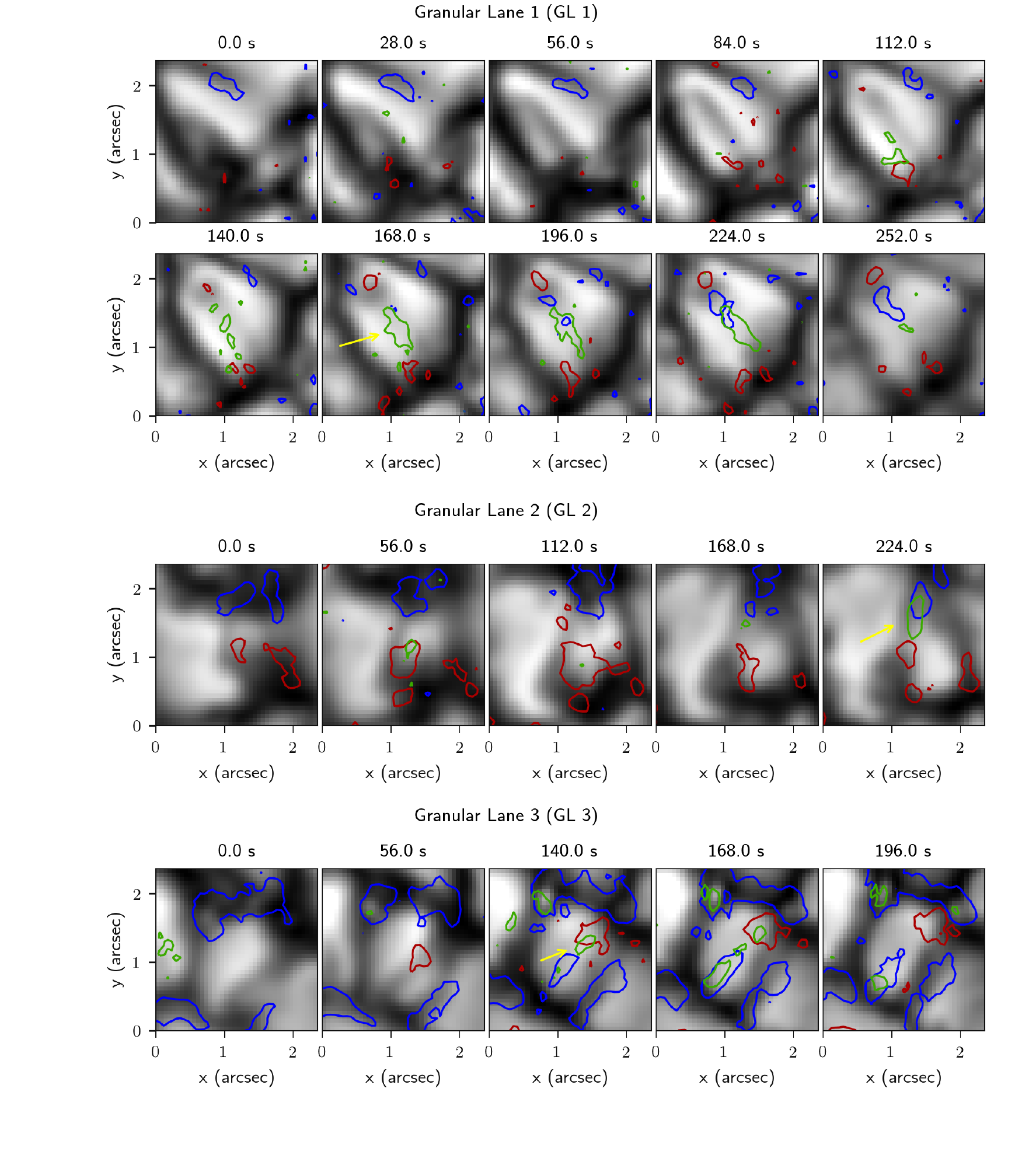}
  \caption{Temporal evolution of three granular lanes observed with the SST. The sequence of maps  shows the continuum intensity \ near the Fe\,{\sc i} 617.3\,nm line normalized to the temporal maximum of the selected FOV for the granular lanes GL1 (top two rows), GL2 (third) row, and GL3 (bottom row). Contours in red and blue mark positive and negative circular polarization at a cutoff level of  $\pm 5 \times 10^{-3} I_c$.  The green contours mark the linear polarization  at a cutoff of $4.7\times 10^{-3} I_c$. Yellow arrows point to the locations where the dark lane of the granular lane is occupied by a substantial patch of linear polarization. An animation of the full temporal evolution is available that includes images with and without polarization contours and runs from $t=-56$ to $280$\,s for all three granular lanes. \label{tempdevelop2}}
        \end{figure*}

The noise in the polarimetric data (Stokes $Q$, $U$ and $V$), calculated as the standard deviation of the signal measured in the continuum of the entire time series, was approximately $\sigma_{Q,U,V}=2.2\times 10^{-3}$ (in units of the spatially and temporally averaged continuum intensity $I_{c}$). We produced maps of the total circular and linear polarization, by averaging over the spectral dimension. The blue/red and green contours in Fig.~\ref{tempdevelop2} correspond to values approximately 10 times larger than the noise in the total negative/positive circular and linear polarization maps, respectively.

The inference of the solar atmospheric parameters (three components of the magnetic field, and line-of-sight velocity) from the polarimetric data was performed via the application of the Very Fast Inversion of the Stokes Vector code \citep[VFISV; ][]{2011SoPh..273..267B}. This code operates under the Milne-Eddington approximation and, as such, provides an average of the atmospheric parameters over the formation region of the spectral line \citep{2014A&A...572A..54B}. The filling factor was set to one, assuming that the magnetic field uniformly occupies a resolution element. 

We used a series of 60 high-resolution CO$^{5}$BOLD simulation snapshots with a cadence of 10\,s. The computational domain spans 9.6\,Mm $\times$ 9.6\,Mm $\times$ 2.8\,Mm, discretized on 960 $\times$ 960 $\times$ 280 grid cells. The bottom boundary was located at $z\approx-1240$~km and the top boundary was at $z\approx1560$~km. The simulation was carried out starting with an initial, homogeneous, vertical field of 50~G magnetic flux density in the entire domain. The first snapshot used in this work was taken after a relaxation phase of 5773\,s in which the magnetic fields quickly had concentrated in the intergranular lanes. 
More details and boundary conditions of the simulation are given in \citet[][Section~2.1]{doi:10.13097/archive-ouverte/unige:115257}, where this particular run is labeled as d3gt57g44v50fc.

We synthesized the full Stokes profiles for the Fe\,{\sc i} 617.3\,nm spectral line using the NICOLE code ~\citep{2015A&A...577A...7S}. To obtain the linear and circular polarization maps, we integrated over the same wavelength range as done with the observations. No degradation of the synthesized profiles was performed for this qualitative comparison with observations.

\section{Results and Discussion} \label{sec:resdis}
Although granular lanes are ubiquitous in the observed continuum images, we find only in a few cases polarization signal associated with them. Figure~\ref{tempdevelop2} shows three cases of significant polarization signal within the granule at the location of the granular lane. In all three cases, the granular lane develops close to an intergranular lane and moves toward the center of the granule (see the animation of Figure~\ref{tempdevelop2}  for the full evolution). The first case (labeled GL 1, top two rows) shows an averaged-sized granule with a granular lane developing on its left border making its way into the granule on a diagonal trajectory from the lower left to the upper right. At $t=84$\,s, the dark lane has reached the center of the granular cell. The dark lane is preceded by a bright rim that travels together with the dark lane. A first feeble linear polarization signal appears at  $t=28$\,s (small green contours), and by $t=168$\,s, when the granular lane already starts to dissolve, we see an extended contiguous elongated patch of linear polarization signal (yellow arrow). It is located right at the position of the granular lane and is elongated in the direction of the lane. The elongated patch of linear polarization stays for approximately 1 minute in time in the later stage of the GL development when the dark lane slowly fades and is not seen anymore in the continuum image by the time $t=224$\,s.

 \begin{figure*}
 \centering
  \includegraphics[,width=\textwidth]{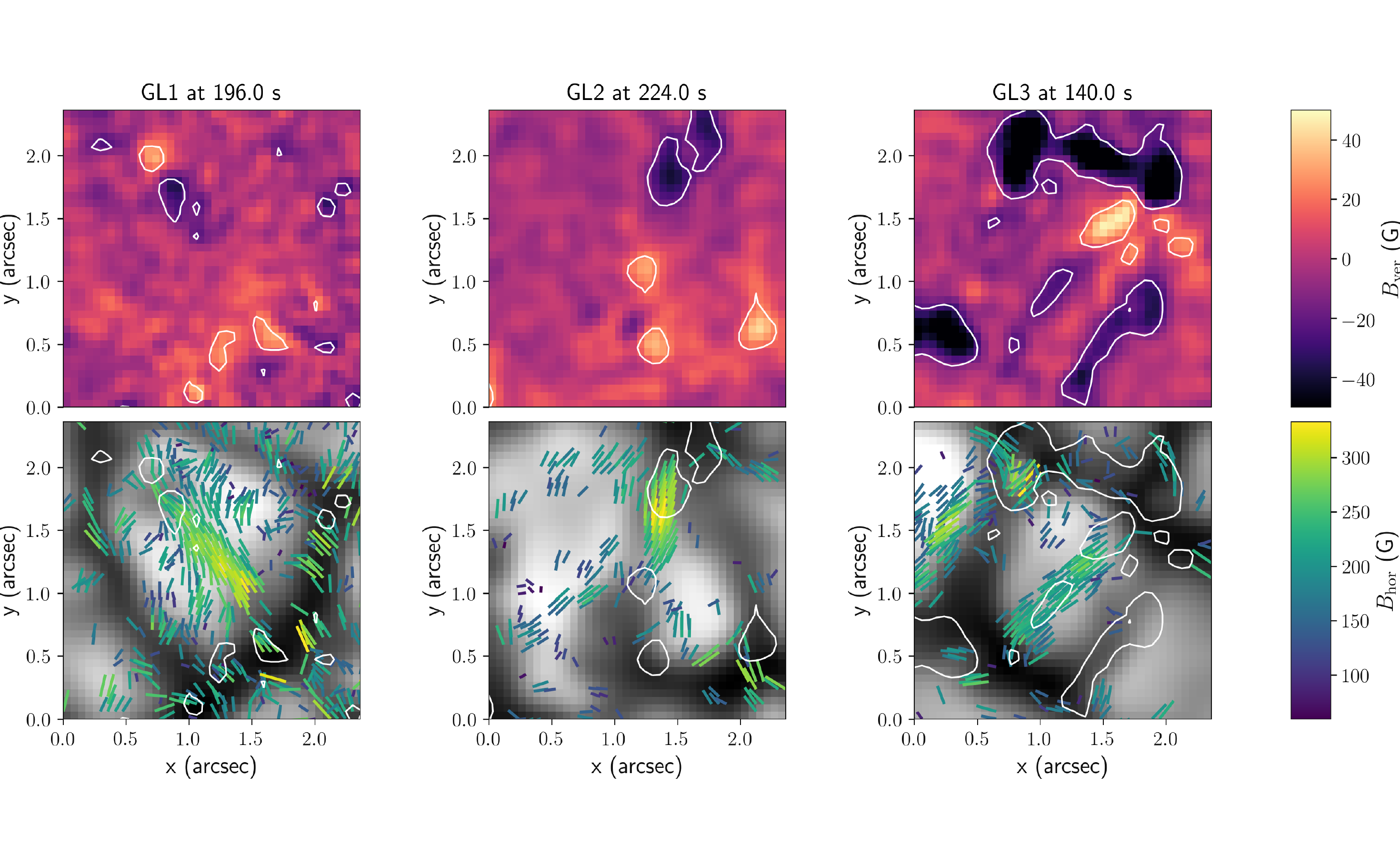}
  \caption{Magnetic field vector information for the three granular lanes GL1, GL2, and GL3 of Fig.~\ref{tempdevelop2}. {\em  Top row:}  vertical magnetic field strength for one time instant of Fig.~\ref{tempdevelop2} for each case.  {\em  Bottom row:}  continuum intensity corresponding to the time instants of the top row. The streaks indicate the direction of the horizontal magnetic field  as derived with the VFISV inversion. This is done only at those locations where either Stokes $Q$ or $U$ is larger than 3 times the noise level. The streaks are color coded corresponding to the total horizontal magnetic field strength. The color bars to the right indicate the vertical and horizontal magnetic field values. White contours mark the regions where $|{B_{\rm ver}}| > 20$\,G. 
  }
   \label{baz}
\end{figure*}

We observe a similar scenario with GL2 and GL3 also shown in Fig.~\ref{tempdevelop2}. GL2 starts at the right granule boundary moving to the left, and GL3 travels from the lower right diagonally toward the upper left corner. In each case, the polarization develops at a later stage of the granular lane evolution when the dark lane already fades. Here too, we find elongated linear polarization patches aligned with the direction of the dark lane (yellow arrows) and flanked by circular polarization patches of opposite polarity at the two ends of the dark granular lane. 
In both cases the surrounding intergranular lanes already harbored some preexisting circular polarization signal, probably not connected to the GL. The opposite polarity patches together with the elongated linear polarization patch connecting them bears a resemblance to a small-scale emerging loop~\citep[e.g.,][]{2007ApJ...666L.137C}) as can be best appreciated in the case of GL2 at time $t=224$\,s. However, similar to the case described by~\cite{2012arXiv1207.6418Y}, we do not observe the linear polarization appearing first, as one would expect from emerging magnetic loops. The same goes for GL3 where the emerging flux shows a more complex configuration.
 \begin{figure*}
 \centering
  \includegraphics[width=\textwidth]{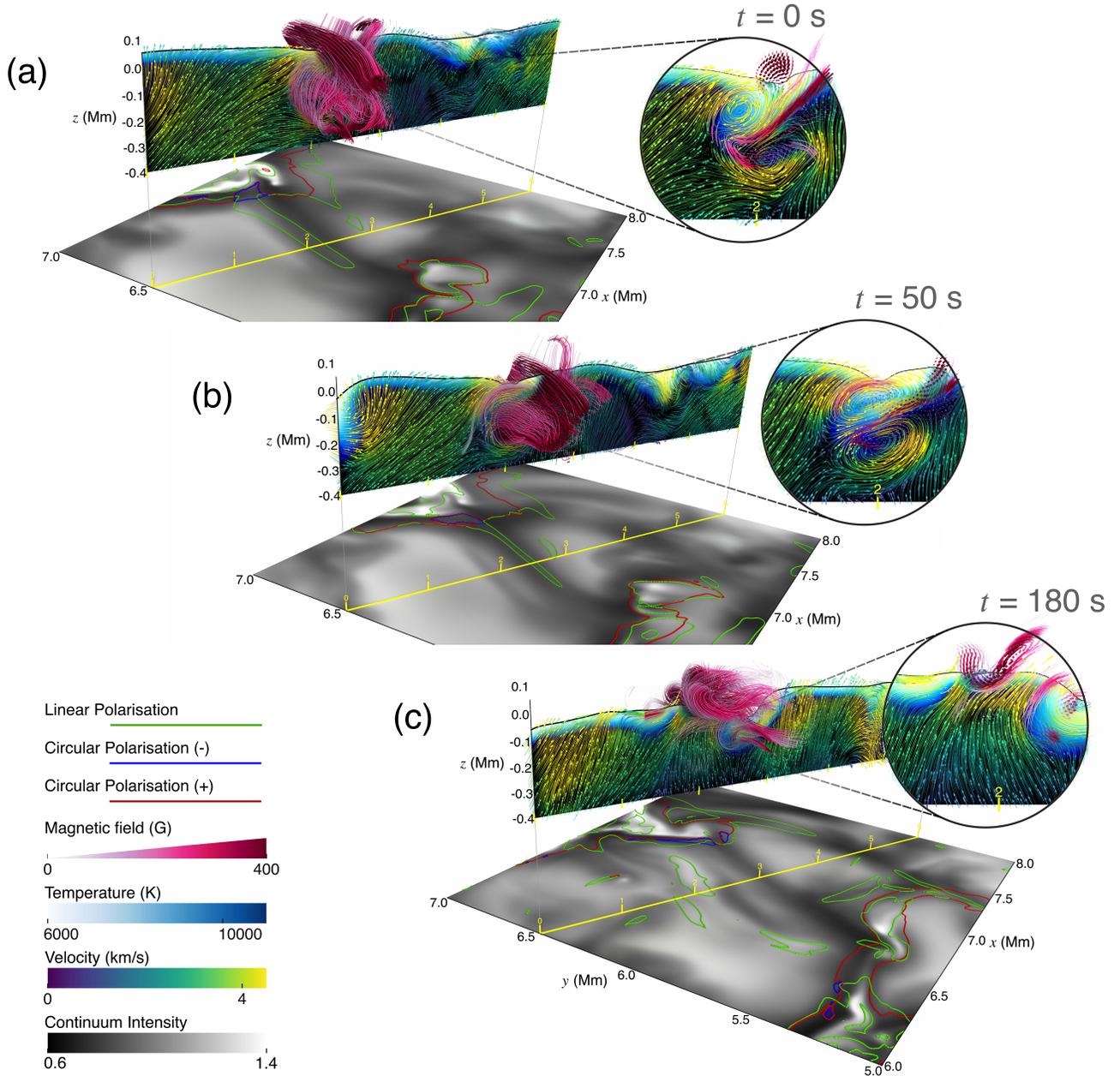}
  \caption{Evolution of a granular lane and associated magnetic field of a CO$^{5}$BOLD simulation. The three panels correspond to time instants taken at (a) $t=0$\,s, (b) 50\,s, and  (c) 180\,s. The bottom scene of each time instant shows in grayscale the continuum intensity near the synthesized spectral line Fe\,{\sc i} 617.3\,nm. The 3D scene above displays the temperature and the flow field below the Rosseland optical depth $\tau_{R}=1$  (black curve).  The location of the section is indicated with a yellow line on the continuum map with vertical markers as a guide to the corresponding point at $z=-0.4$~Mm in the 3D scene above. The contours in green outline the linear polarization calculated from the synthesized Stokes profiles at a threshold of  $1.0\times 10^{-3} I_c$. The  circular polarization is marked by red and blue contours. The magnetic field lines that cross the section plane are shown in pink color. The circular close-up view at each time instant provides a clear look at the temperature, flow field, and magnetic field orientation at the location of the granular lane. Here, $t=0$\,s refers to the simulation time $t_{\rm sim}=6210$\,s since the start of the actual simulation d3gt57g44v50fc. The continuum intensity is normalized with the intensity of the HSRA reference atmosphere at the same wavelength. An animation of this figure is available, which includes two simulation cases, running from $t=0$ to $180$\,s and $150$\,s, respectively. For each simulation the animation loops twice.}
   \label{sim}
\end{figure*}
Figure~\ref{baz} relays the magnetic vector information from Stokes inversion for the three cases, whereby we show one time instance of each of the time series shown in Fig.~\ref{tempdevelop2}. 
In all three cases the longitudinal magnetic field strength (top row) is rather weak with values ranging from 20 to 50\,G. The only exception is the negative polarity region in GL3 that reaches  field strengths up to 110\,G. The bottom row of Fig.~\ref{baz} shows the direction of the horizontal magnetic field. Following \citet{2012A&A...547A..89B}, we plot the horizontal magnetic field only at those locations where the signal in either Stokes $Q$ or $U$ is larger than 3 times the noise level. In all three cases, the horizontal magnetic field is aligned with the dark granular lane reaching a field strength ($B_{\rm hor}$) of up to 300\,G. Hence, the horizontal magnetic field points toward the location of longitudinal magnetic field flanking the linear polarization. This finding may come as a surprise as one could intuitively expect the vortex tube to wrap the magnetic field azimuthally around the vortex axis so that the magnetic field would be directed perpendicular to the dark granular lane instead of parallel to it. To clarify this point we turn to simulations for further insights.

We identify several granular lanes in the CO$^{5}$BOLD simulation that show structural similarities with observations. In Fig.~\ref{sim} we look at a typical example. An animation is provided that gives a more detailed view on the evolution of the granular lane of Figure~\ref{sim} and includes an additional example.
Figure~\ref{sim} shows three representative snapshots during the lifetime of a granular lane. 

In Fig.~\ref{sim}(a), we see in the continuum image (grayscale map) a dark intergranular lane oriented diagonally from the upper left to the lower right running across the image center. It exhibits significant linear polarization (green contours) almost throughout the entire length of the lane. The field lines in this location, plotted in pink color in the 3D  scene above, reveal a strong horizontal magnetic flux bundle aligned with the intergranular lane. A significant section of it is located above the Rosseland optical depth $\tau_{R}=1$ layer (black curve) where the linear polarization signal forms, whereas the rest of the flux bundle dips below the $\tau_{R}=1$  surface. 
The field lines are connected on both ends (not shown in the 3D  scene) to vertical flux concentrations that can be seen as circular polarization patches located on opposite sides of the intergranular lane in the synthesized images (red contour in the lower right and blue contour in the upper left part of the intergranular lane). The velocity field (indicated by arrows) shows a strong vortical flow near the $\tau_{R}=1$ layer with the flow field reaching into the intergranular lane (see the close-up inset). This is a typical feature of horizontal vortex tubes described in \cite{2010ApJ...723L.180S}. An oppositely oriented vortex flow can be seen deeper down, presumably a result of angular momentum conservation. After 50\,s, in Fig.~\ref{sim}(b), we see in the continuum image the characteristic dark lane of a granular lane oriented perpendicular to and moving along the yellow straight line at position 2 toward the center of the granule. The dip in the $\tau_{R}=1$ surface corresponding to this lane and its position relative to the horizontal field of the intergranular lane can be seen again in the 3D scene above as well as in the corresponding close-up inset. The horizontal magnetic field lines, that were earlier occupying the intergranular space, are dragged to the deeper layers by the strong flow present in between the two vortices. This reduces the magnetic flux in the intergranular lane, as also evident from the decrease in the linear polarization signal.  In Fig.~\ref{sim}(c), 180\,s later and already in the decaying phase of the granular lane, we see the reemergence of the field lines that were grabbed by the flow field from the interganular lane. These fields predominantly retain their strength and orientation during the reshuffling but become also twisted to form a (weaker) component in the direction of the fluid flow, i.e, perpendicular to the granular lane. We notice islands of linear polarization signal in the granular lane at this point in time corresponding mainly to the strong lane-aligned field component emerging from the surface of $\tau_{R}=1$. At close inspection of the simulated atmosphere we note that in this case the lane-aligned component (the cross-section of these field lines appear as dots in the close-up view) rises little above the $\tau_{R}=1$ layer, barely reaching the formation height of the chosen Fe\,{\sc i} line, but still leading to a weak linear polarization signal in the synthesized spectrum. Its strength of approximately 400\,G at $\tau_{R}=1$ is in rough agreement with our results shown in Fig.~\ref{baz}.

Another simulation case,  included in the animation of Figure~\ref{sim},  exhibits the same qualitative behavior, but shows a slightly more complex configuration. A horizontal vortex tube is initiated  at the boundary of a granule. Preexisting horizontal fields are trapped within the vortex and transported into the granule. In this second case, however, the initial horizontal field within the intergranular lane lies below the $\tau_{R}=1$ layer (around $z\approx-0.25$\,Mm) and therefore escapes polarimetric detection. This is similar to our observations in which we do not observe the initial horizontal field in the intergranular space. Again, the initial horizontal field is then brought to the surface by the vortical flow associated with the granular lane. Linear polarization in intergranular lanes has been reported by \cite{2010ApJ...723L.149D} who find 8\% of the detected linear polarization patches being embedded over an extended time period in the downflows of intergranular lanes, traditionally thought to be occupied by vertical magnetic field. Also in this second simulation case (in the animation of Figure~\ref{sim}), there exists in addition to the emerged magnetic field that is aligned with the vortex-tube axis a weaker toroidal component perpendicular to it. However, along the chosen section, the aligned component is  significantly stronger than the toroidal one. We also see a much stronger synthesized polarization signal compared to the first simulation case in Fig.~\ref{sim}. The two examples showcase the way in which the observed polarimetric signal can be produced, although each case differs in the details of these signals.

The reason we find only a few examples of granular lanes harboring a polarization signal may have various causes. One being that the horizontal magnetic field that is recycled to the surface is too weak or does not reach the formation region of the spectral line that was used in our observations. Another reason may be that the initial magnetic field being grabbed by the vortex flow in the intergranular space is less coherent than the one shown in Fig.~\ref{sim} and is for example of a more turbulent nature. In this case, the reemerging magnetic field is expected to retain its turbulent nature and would therefore escape detection based on the Zeeman effect due to polarimetric cancellation.

In the small-scale dynamo simulation analyzed by \cite{2018ApJ...859..161R}, granular lanes were launched from the interior of  exploding granules next to the newly forming downflow in the granule center. He sees indeed turbulent magnetic fields in the downflow lane being transported by the vortex flow of the granular lane back to the neighboring granular upflow, where it reappears as a turbulent field. In any case, the  basic underlying process of magnetic field being  transported and brought back to the surface by a vortical flow remains the same in both the simulation of \cite{2018ApJ...859..161R} and our observations and simulations of the present Letter. \cite{2018ApJ...859..161R} refers to this process as shallow (local) recycling of magnetic fields and shows that it is associated with substantial field amplification.

\section{Summary and Conclusion} \label{sec:concl}

We have found clear observational evidence of magnetic field associated with the granular lanes initially identified by ~\cite{2010ApJ...723L.180S} as vortex tubes. Our findings can be summarized as follows:
\begin{enumerate}
  \item The observed polarization develops at the late stage of the visible dark granular lane traveling into the granule. As the dark lane fades, we observe linear polarization that extends parallel to the granular lane and, either at the same time or prior to it, circular polarization patches of opposite polarity at either end of the granular lane.
  \item The magnetic field shows a loop-like structure with the horizontal magnetic field component being aligned with the dark lane and having a strength of several 100\,G. The observed magnetic field is comparable in strength to that found in small-scale emerging magnetic loops as in \cite{2010ApJ...714L..94M} (approx. 200 G). Weaker (around 50G) vertical fields are observed at both ends of the granular lane. 
  \item Simulations show how preexisting horizontal magnetic field residing in the intergranular lanes, either at or below the surface, can be trapped in horizontal vortex tubes developing at granule boundaries. These horizontal fields are consequently transported through the vortex flow into the granule and upward, finally appearing at the granule surface.
\end{enumerate}
We have witnessed the emergence of magnetic fields in granule interiors, obviously brought to the surface by the plasma flow associated with granular lanes. 
Since granular lanes have been identified with vortical flows of subgranular scale, this magnetic flux was likely transported by this flow from the integranular space to the granule interior in a shallow, recycling fashion. This scenario is suggested by our simulations that show remarkably similar polarimetric signals and continuum intensity features as is observed. Small-scale dynamo simulations of \cite{2018ApJ...859..161R} have shown a similar process to exist in exploding granules, in which a shear-flow-amplified magnetic field is transported to the surface through shallow recirculation of magnetic fields by the vortical flow of granular lanes. Accordingly, this shallow recirculation is a significant source of magnetic field for the small-scale turbulent dynamo that is supposed to be at the origin of the quiet-Sun magnetic field. 

On the way from the intergranular lane to the granule interior, the magnetic field will likely be stretched and thus amplified.
Ultimately, some of this field becomes twisted, as is suggested by the simulations, and transported by the granular flow back to the intergranular lane where it is folded onto the preexistent intergranular magnetic field. Such a complex magnetic field and flow topology, which is a basic condition for the working of the turbulent dynamo, can in principle be searched for making use of polarimetry with multiple spectral lines. Here, we have found the first hint of it. Therefore, the present observations offer a first glimpse on the workings of the small-scale dynamo on the Sun.
Observations with the newly commissioned Daniel K. Inouye Solar Telescope~\citep[DKIST; e.g.][]{2016AN....337.1064T} solar telescope, targeting high polarimetric sensitivity and deep-seated magnetic fields, possibly combining Zeeman with Hanle measurements, should provide new insights into the interaction of vortical flows with the magnetic field embedded in the turbulent plasma. It would also provide new insights into the workings of the small-scale dynamo and the origin of the quiet-Sun magnetic field.

\acknowledgments

\noindent The Swedish 1 m Solar Telescope is operated on the island of La Palma by the Institute for Solar Physics of Stockholm University in the Spanish Observatorio del Roque de los Muchachos of the Instituto de Astrof\'{i}sica de Canarias. The Institute for Solar Physics is supported by a grant for research infrastructures of national importance from the Swedish Research Council (registration number 2017-00625). This project has received funding from the European Union's Horizon 2020 research and innovation program under grant agreement No. 824135, the Trans-National Access Programme of SOLARNET. We also thank Oleksii Andriienko for performing the data reconstruction with SSTRED. C.E.F. is funded by the Leibniz Association grant for the SAW-2018-KIS-2-QUEST project. G.V. thanks funds from the ESCAPE project of the European Union’s Horizon 2020 Research and Innovation Programme (Grant Agreement No. 824064). F.C. is supported through the CHROMATIC project (2016.0019) funded by the Knut and Alice Wallenberg foundation. Simulations were carried out at the Swiss National Supercomputing Centre (CSCS) under project ID s560 with support from the Swiss National Science Foundation under grant ID 200020\_157103.
We are grateful to Wolfgang Schmidt for his valuable comments on the manuscript and greatly appreciate the constructive comments by an anonymous referee, which both helped to significantly improve the manuscript.
\bibliography{cef_manuscript_gl}{}
\bibliographystyle{aasjournal}

\end{document}